\documentstyle[12pt]{article}

\begin{document}

\begin{titlepage}
\setcounter{page}{1}
\title{\bf Relativistically Extended Modification of 
the Schr\"{o}dinger Equation}
\author{Waldemar Puszkarz\thanks{Electronic address: puszkarz@cosm.sc.edu}
\\
\small{\it  Department of Physics and Astronomy,}
\\
\small{\it University of South Carolina,}
\\
\small{\it Columbia, SC 29208}}
\date{\small (May 15, 1999)}
\maketitle
\begin{abstract}
We propose a nonlinear modification of the Schr\"{o}dinger 
equation that possesses the main properties of this equation such as
the Galilean invariance, the weak separability of composite systems, 
and the homogeneity in the wave function. The modification is derived from 
the relativistic relation between the energy and momentum of free particle 
and, as such, it is the best relativistic extension of the Schr\"{o}dinger 
equation that preserves the properties in question. The only change it 
effectively entails in the Schr\"{o}dinger equation involves the conserved 
probability current. It is pointed out that it partially retains the linear 
superposition principle and that it can be used to model the process of 
decoherence.

\vskip 0.5cm
\noindent
\end{abstract}
\end{titlepage}

Since its discovery in 1926, the Schr\"{o}dinger equation, 
\begin{equation}
i\hbar \frac{\partial \Psi }{\partial t}=-\frac{\hbar ^{2}}{2m}\Delta \Psi
+V\Psi ,  \label{1}
\end{equation}
has been subjected to a very thorough examination, both experimental and
theoretical. In the very same year, Madelung showed that one can represent
this equation as a system of two nonlinear equations, 
\begin{equation}
\hbar \frac{\partial R^{2}}{\partial t}+\frac{\hbar ^{2}}{m}\vec{\nabla}%
\cdot \left( R^{2}\vec{\nabla}S\right) =0,  \label{2}
\end{equation}
\begin{equation}
\frac{\hbar ^{2}}{m}\Delta R-\frac{\hbar ^{2}}{m}R\left( \vec{\nabla}%
S\right) ^{2}-2\hbar R\frac{\partial S}{\partial t}-2RV=0,  \label{3}
\end{equation}
for the amplitude $R$ and the phase $S$ of the wave function $\Psi =Re^{iS}$%
. They form the so-called hydrodynamic representation of the Schr\"{o}dinger
equation and can be obtained as the imaginary and real part of this
equation, correspondingly. The first of these equations is the continuity
equation with the probability current 
\begin{equation}
\vec{j}_{Sch}=\frac{\hbar ^{2}}{m}R^{2}\vec{\nabla}S  \label{4}
\end{equation}
and the probability density $\rho =R^{2}$. However, it is the other equation
that is crucial for determining the stationary states of quantum-mechnical
systems defined by the condition $\partial R^{2}/\partial t=0$.

Let us recall that, heuristically, the free Schr\"{o}dinger equation can be
derived from the relativistic dispersion relation between the energy and
momentum of a free particle 
\begin{equation}
E=c\sqrt{p^{2}+\left( mc\right) ^{2}}=mc^{2}\sqrt{1+\left( \frac{p}{mc}%
\right) ^{2}}.  \label{5}
\end{equation}
One obtains it by discarding the rest energy term from the leading
approximation to (5), 
\begin{equation}
E=mc^{2}+\frac{p^{2}}{2m},  \label{6}
\end{equation}
and applying the first quantization procedure. To account for a wider range
of physical situations, one can subsequently add the potential term to the
RHS of (6). This leads directly to (1).

It is the purpose of this paper to derive a relativistic extension of the
Schr\"{o}dinger equation which would stem from a closer approximation to (5)
than the truncated first order approximation (6) and which, in fact, would
be a nonlinear modification of the equation concerned.\footnote{%
For a broader exposition of various aspects of nonlinear quantum mechanics
see \cite{Pusz1}. More elaborate, though still far from comprehensive a list
of approaches and motivations to modify this equation can be found in \cite
{Gold}.}

The Schr\"{o}dinger equation is homogeneous of degree one in the wave
function, however it is not necessarily clear if this property is underlied
by some important physical principle or whether it entails any physically
relevant consequences. It is thus conceivable that this property is an
accidental feature of the Schr\"{o}dinger equation and, thus, disposing of
it in a nonlinear modification of this equation will most likely not cause
any  damage. Nevertheless, as pointed out in \cite{Pusz1}, nonhomogeneous
modifications of the Schr\"{o}dinger equation suffer from the ambiguity in
the definition of the energy functional. Although this occurs also in
homogeneous nonlinear variants of this equation, there exists a class of
homogeneous modifications for which the quantum-mechanical energy functional
defined as the expectation value of the corresponding Hamiltonian operator
coincides with the energy defined as a conserved quantity within the
Lagrangian framework \cite{Pusz1}. This indicates that the homogeneity is
necessary but not sufficient to guarantee the uniqueness of the energy
functional. Weinberg has elevated the discussed property to one of the
fundamental assumptions of his original generalization of the linear
framework of quantum mechanics \cite{Wein1, Wein2}. It was his hope,
particularly expressed in \cite{Wein2}, that the homogeneity of nonlinear
variants of this equation might entail their separability.\footnote{%
As stated in \cite{Wein2}: ``The problem of dealing with separated systems
has led other authors to limit possible nonlinear terms in the
Schr\"{o}dinger equation to a logarithmic form;'' and ``The homogeneity
assumption (2) makes this unnecessary.''} Even if this is true in his
modification and was shown to be valid in some others \cite{Gold, Pusz2},
one can find examples of nonlinear variants of the Schr\"{o}dinger equation
that contradict this thesis \cite{Pusz3}. Yet, it should be noted \cite
{Pusz3, Pusz4} that the homogeneity is essential for nonlinear modifications
of this equation to possess a weakly separable multi-particle extension.

The separability of noninteracting subsystems \cite{Bial, Gold} has much
more profound and better understood physical implications. To cast more
light on this issue, let us demonstrate the weak separability of the
Schr\"{o}dinger equation in the hydrodynamic formulation. We are considering
a quantum system made up of two noninteracting subsystems in the sense that 
\cite{Bial} 
\begin{equation}
V(\vec{x}_{1},\vec{x}_{2},t)=V_{1}(\vec{x}_{1},t)+V_{2}(\vec{x}_{2},t).
\label{7}
\end{equation}
We will show that a solution of the Schr\"{o}dinger equation for this system
can be put in the form of the product of wave functions for individual
subsystems for any $t>0$, that is, $\Psi (x_{1},x_{2},t)=\Psi
_{1}(x_{1},t)\Psi _{2}(x_{2},t)=R_{1}(x_{1},t)R_{2}(x_{2},t)exp\left\{
i(S_{1}(x_{1},t)+S_{2}(x_{2},t))\right\} $ and that this form entails the
separability of the subsystems. The essential element here is that the
subsystems are initially uncorrelated which is expressed by the fact that
the total wave function is the product of $\Psi _{1}(\vec{x}_{1},t)$ and $%
\Psi _{2}(\vec{x}_{2},t)$ at $t=0$. What we will show then is that the
subsystems remain uncorrelated during the evolution and that, at the same
time, they also remain separated. It is the additive form of the total
potential that guarantees that no interaction between the subsystems occurs,
ensuring that they remain uncorrelated during the evolution. However, such
an interaction may, in principle, occur in nonlinear modifications of the
Schr\"{o}dinger equation even if the form of the potential itself does not
imply that. This is due to a coupling that a nonlinear term usually causes
between $\Psi _{1}(\vec{x}_{1},t)$ and $\Psi _{2}(\vec{x}_{2},t)$. As a
result, even in the absence of forces the very existence of one of the
particles affects the evolution of the other one, clearly violating
causality. This particular feature of the Schr\"{o}dinger equation
distinguishes it from the Klein-Gordon equation. One arrives at the latter
from (5) upon squaring it and employing the first quantization. The time
derivative of this equation introduces a coupling between $\Psi _{1}(\vec{x}%
_{1},t)$ and $\Psi _{2}(\vec{x}_{2},t)$, rendering the equation inseparable
in the sense discussed. Another physically significant difference between
these equations is in the probability density which can be negative for the
Klein-Gordon equation.

The discussed separability is called the weak separability since it assumes
that the wave function of the total system is the product of the wave
functions of its subsystems in contradistinction to the strong version of
separability that does not employ this assumption. As shown by L\"{u}cke 
\cite{Luc1, Luc2}, weakly separable modifications, such as the modification
of Bia\l ynicki-Birula \cite{Bial} or the Doebner-Goldin modification, \cite
{Doeb1} can still violate separability when the compound wave function is
not factorizable, and thus they are not strongly separable. An alternative
effective approach to the strong separability has been proposed by Czachor 
\cite{Czach1}. This approach treats the density matrix as the basic object
subjected to the quantum equations of motion which are modified\footnote{%
What this means in practice is that the basic equation is the nonlinear von
Neumann equation \cite{Czach2} for the density matrix instead of some
nonlinear Schr\"{o}dinger equation for the pure state.} compared to a
nonlinear Schr\"{o}dinger equation for the pure state. It admits a large
class of nonlinear modifications including those ruled out by the
fundamentalist approach advocated by L\"{u}cke and even those that are not
weakly separable as, for instance, the cubic nonlinear Schr\"{o}dinger
equation.

The Schr\"{o}dinger equation for the total system, assuming that the
subsystems have the same mass $m$, reads now

\begin{eqnarray}
\lefteqn{\hbar \frac{\partial R_{1}^{2}R_{2}^{2}}{\partial t}+\frac{\hbar
^{2}}{m}\left\{ \left( \vec{\nabla}_{1}+\vec{\nabla}_{2}\right) \cdot \left[
R_{1}^{2}R_{2}^{2}\left( \vec{\nabla}_{1}S_{1}+\vec{\nabla}_{2}S_{2}\right)
\right] \right\} = }  \nonumber \\
&&\hbar R_{2}^{2}\frac{\partial R_{1}^{2}}{\partial t}+\hbar R_{1}^{2}\frac{%
\partial R_{2}^{2}}{\partial t}+\frac{\hbar ^{2}}{m}R_{2}^{2}\vec{\nabla}%
_{1}\cdot \left( R_{1}^{2}\vec{\nabla}_{1}S_{1}\right) +\frac{\hbar ^{2}}{m}%
R_{1}^{2}\vec{\nabla}_{2}\cdot \left( R_{2}^{2}\vec{\nabla}_{2}S_{2}\right) =
\nonumber \\
&&R_{1}^{2}R_{2}^{2}\hbar \left\{ \frac{1}{R_{1}^{2}}\frac{\partial R_{1}^{2}%
}{\partial t}+\frac{\hbar ^{2}}{m}\frac{1}{R_{1}^{2}}\vec{\nabla}_{1}\cdot
\left( R_{1}^{2}\vec{\nabla}_{1}S_{1}\right) +\left[ \hbar \frac{1}{R_{2}^{2}%
}\frac{\partial R_{2}^{2}}{\partial t}++\frac{\hbar ^{2}}{m}\frac{1}{%
R_{2}^{2}}\vec{\nabla}_{2}\cdot \left( R_{2}^{2}\vec{\nabla}_{2}S_{2}\right)
\right] \right\} =0  \label{8}
\end{eqnarray}

and

\begin{eqnarray}
\lefteqn{\frac{\hbar ^{2}}{m}\left( \Delta _{1}+\Delta _{2}\right)
R_{1}R_{2}-2\hbar R_{1}R_{2}\frac{\partial (S_{1}+S_{2})}{\partial t}-\frac{%
\hbar ^{2}}{m}R_{1}R_{2}\left( \vec{\nabla}_{1}S_{1}+\vec{\nabla}%
_{2}S_{2}\right) ^{2}-}  \nonumber \\
&&\left( V_{1}+V_{2}\right) R_{1}R_{2}=\frac{\hbar ^{2}}{m}R_{2}\Delta
_{1}R_{1}+\frac{\hbar ^{2}}{m}R_{1}\Delta _{2}R_{2}-2\hbar R_{1}R_{2}\frac{%
\partial S_{1}}{\partial t}-2\hbar R_{1}R_{2}\frac{\partial S_{2}}{\partial t%
}  \nonumber \\
&&+\frac{\hbar ^{2}}{m}R_{1}R_{2}\left( \vec{\nabla}_{1}S_{1}\right) ^{2}+%
\frac{\hbar ^{2}}{m}R_{1}R_{2}\left( \vec{\nabla}_{2}S_{2}\right)
^{2}-V_{1}R_{1}R_{2}-V_{2}R_{1}R_{2}=  \nonumber \\
&&R_{1}R_{2}\left\{ \left[ \frac{\hbar ^{2}}{m}\frac{\Delta _{1}R_{1}}{R_{1}}%
-2\hbar \frac{\partial S_{1}}{\partial t}+\frac{\hbar ^{2}}{m}\left( \vec{%
\nabla}_{1}S_{1}\right) ^{2}-V_{1}\right] +\left[ \frac{\hbar ^{2}}{m}\frac{%
\Delta _{2}R_{2}}{R_{2}}-2\hbar \frac{\partial S_{2}}{\partial t}+\right.
\right.  \nonumber \\
&&\left. \left. \frac{\hbar ^{2}}{m}\left( \vec{\nabla}_{2}S_{2}\right)
^{2}-V_{2}\right] \right\} =0.  \label{9}
\end{eqnarray}
Implicit in the derivation of these equations is the fact that $\vec{\nabla}%
_{1}f_{1}\cdot \vec{\nabla}_{2}g_{2}=0$, where $f_{1}$ and $g_{2}$ are
certain scalar functions defined on the configuration space of particle 1
and 2, correspondingly. What we have obtained is a system of two equations,
each consisting of terms (in square brackets) that pertain to only one of
the subsystems. By dividing the first equation by $R_{1}^{2}R_{2}^{2}$ and
the second one by $R_{1}R_{2}$, one completes the separation of the
Schr\"{o}dinger equation for the compound system into the equations for the
subsystems. Moreover, we have also showed that indeed the product of wave
functions of the subsystems evolves as the wave function of the total system.

We would like the modified Schr\"{o}dinger equation to possess the property
of separability of composite noninteracting systems in the sense described
and to be Galilean invariant. It should also ensure a non-negative
probability density emerging in the continuity equation. Moreover, we
require that the probability current of this equation be conserved. These
properties are certainly of physical relevance. As we will see, our approach
will result in a homogeneous equation by way of construction, exactly as the
Schr\"{o}dinger equation is rendered homogenous when derived in the manner
outlined above.

To proceed, let us note that by expanding (5) in a series of $p^{2}$ and
applying the first quantization we obtain terms $\lambda _{c}^{2n}\Delta
^{n}\Psi $, where $\lambda _{c}=\frac{\hbar }{mc}$ is the Compton
wavelength. Employing the Schr\"{o}dinger-Madelung representation, we
observe that the terms concerned contain growing with $n$ a number of terms
involving derivatives of $R$ and $S$. The essence of our approach is to
extract from them the terms that would satisfy our criteria. It is easy to
check that the following reccurence relation emerges for $n\ge 1$ 
\begin{eqnarray}
\Delta ^{n}\Psi &=&(A_{n}+iB_{n})\exp (iS)=\left\{ \left[ (\Delta -(\vec{%
\nabla}S)^{2})A_{n-1}-(\Delta S+2\vec{\nabla}S\cdot \vec{\nabla}%
)B_{n-1}\right] +\right.  \nonumber \\
&&\left. i\left[ (\Delta -(\vec{\nabla}S)^{2})B_{n-1}+(\Delta S+2\vec{\nabla}%
S\cdot \vec{\nabla})A_{n-1}\right] \right\} \exp (iS),  \label{10}
\end{eqnarray}
where $A_{0}=R$ and $B_{0}=0$. The first part of this equality is trivially
valid for $n=0$ as well. It should be noted that by expanding (5) in a
series of $\Delta ^{n}\Psi $ one obtains nonseparable terms, the only
exception being the leading $n=1$ term that gives rise to the
Schr\"{o}dinger equation. Moreover, such an expansion would not be Galilean
invariant. To ensure that both of these properties are maintained in our
modification, it is more convenient if not necessary to work in the
Schr\"{o}dinger-Madelung representation. To facilitate more general
considerations, let us first examine in more detail the lowest order terms
emerging from (10). These are 
\begin{eqnarray}
A_{1} &=&\Delta R-(\vec{\nabla}S)^{2}R=\left[ \Delta -(\vec{\nabla}%
S)^{2}\right] R,  \label{11} \\
B_{1} &=&(\Delta S+2\vec{\nabla}S\cdot \vec{\nabla})R=\frac{\vec{\nabla}%
\cdot \left( R^{2}\vec{\nabla}S\right) }{R},  \label{12}
\end{eqnarray}
and 
\begin{eqnarray}
A_{2} &=&\left[ \Delta -(\vec{\nabla}S)^{2}\right] ^{2}R-(\Delta S+2\vec{%
\nabla}S\cdot \vec{\nabla})\frac{\vec{\nabla}\cdot \left( R^{2}\vec{\nabla}%
S\right) }{R},  \label{13} \\
B_{2} &=&\left[ \Delta -(\vec{\nabla}S)^{2}\right] \frac{\vec{\nabla}\cdot
\left( R^{2}\vec{\nabla}S\right) }{R}+\left[ \Delta S+2\vec{\nabla}S\cdot 
\vec{\nabla}\right] \left[ \Delta -(\vec{\nabla}S)^{2}\right] R.  \label{14}
\end{eqnarray}

Among them, only $B_{1}$ and $\Delta B_{1}$ are separable, although not
Galilean invariant. $RB_{1}$ is proportional to the divergence of the
probability current of the Schr\"{o}dinger equation, but the continuity
equation (2) is Galilean invariant only as a result of cancellation of
noninvariant contributions coming from its both terms. To see this, let us
recall that under the Galilean transformation, 
\begin{equation}
t=t^{\prime },\,\,\,\vec{x}=\vec{x}^{\prime }+\vec{v}t,  \label{15}
\end{equation}
the amplitude of wave function does not change, the phase changes according
to 
\begin{equation}
S(t,\vec{x})=S^{^{\prime }}(t^{\prime },\vec{x}^{\prime })+m\vec{v}\cdot 
\vec{x}^{\prime }+\frac{m}{2}\vec{v}^{2}t^{\prime },  \label{16}
\end{equation}
whereas the operators that affect these quantities in (2-3) transform as 
\begin{equation}
\vec{\nabla}=\vec{\nabla}^{^{\prime }},\,\,\,\frac{\partial }{\partial t}=%
\frac{\partial }{\partial t^{^{\prime }}}-\vec{v}\cdot \vec{\nabla}%
^{^{\prime }}.  \label{17}
\end{equation}

As one can deduce from the general reccurence rule (10) and the last two
equations, only the terms $\Delta A_{n}$ and $\Delta B_{n}$ could produce
Galilean invariant and separable terms from the terms that became separable
in a preceding iteration, but were not necessarily Galilean invariant. The
conditions of separability and Galilean invariance are stringent enough to
select only one term that maintains separability in successive iterations of 
$\Delta $ and at the same time yields terms that are Galilean invariant. It
is 
\begin{equation}
\alpha =B_{1}=R\Delta S+2\vec{\nabla}S\cdot \vec{\nabla}R.  \label{18}
\end{equation}
Let us note that even if $\vec{\nabla}S\cdot \vec{\nabla}R$ is separable, it
is no longer so when acted upon by $R\Delta $ as it gives rise to
nonseparable terms 
\begin{equation}
R\Delta (\vec{\nabla}S\cdot \vec{\nabla}R)=\sum_{i}R\vec{\nabla}\cdot \left[
\partial _{i}S\vec{\nabla}\cdot \left( \partial _{i}R\right) +\partial _{i}R%
\vec{\nabla}\cdot \left( \partial _{i}S\right) \right] .  \label{19}
\end{equation}
The same applies to other terms of the form $R\Delta \left( \vec{\nabla}%
\Delta ^{2n}S\cdot \vec{\nabla}R\right) $ ($n\geq 1$) that emerge in higher
iterations. For this reason they will be discarded. Now, with the second
term of (18) discarded, $R\Delta \alpha =\vec{\nabla}\cdot (R^{2}\vec{\nabla}%
^{3}S)+R\Delta R\Delta S$, but since $R\Delta R\Delta S$ is not separable,
it should be excluded from our construction as should other terms of the
form $R\Delta R\Delta ^{n}S$ ($n\geq 1$) appearing in subsequent iterations.
Repeating this procedure for $R\Delta ^{n}\alpha $ ($n\ge 1$) one observes
that it effectively produces $\vec{\nabla}\cdot (R^{2}\vec{\nabla}\Delta
^{n}S)$.

The multiplication of all terms of the expansion in question by $R$ is
required in order to arrive at the correct form of the continuity equation.
For the term $B_{1}$ this has already been pointed out. Namely, it is $%
RB_{1}=R\alpha $ and not $B_{1}$ itself that gives rise to the divergence of
the probability current. In the same manner, the terms $R\Delta ^{n}\alpha $
($n\ge 1$) contribute to the divergence of a new probability current in the
modified continuity equation. We see that the only change our construction
of the relativistically extended Schr\"{o}dinger equation brings about is in
this current, or, what amounts to the same, in the continuity equation of
the Schr\"{o}dinger-Madelung system (2-3). One can write this equation as 
\begin{equation}
\hbar \frac{\partial R}{\partial t}=mc^{2}\lambda _{c}^{2}\sum_{n=0}^{\infty
}c_{n}\left( \lambda _{c}^{2}\Delta \right) ^{n}\alpha ,  \label{20}
\end{equation}
where $c_{n}$ are the coefficients of the expansion. Multiplying both sides
of it by $R$ one arrives at 
\begin{equation}
\frac{\partial R^{2}}{\partial t}=\frac{2\hbar }{m}\sum_{n}\lambda
_{c}^{2n}c_{n}R\Delta ^{n}\alpha =\frac{2\hbar }{m}\sum_{n=0}^{\infty
}\lambda _{c}^{2n}c_{n}\vec{\nabla}\cdot (R^{2}\vec{\nabla}\Delta ^{n}S)=%
\frac{2\hbar }{m}\vec{\nabla}\cdot \left[ R^{2}\vec{\nabla}%
\sum_{n=0}^{\infty }\lambda _{c}^{2n}c_{n}\Delta ^{n}S\right] .  \label{21}
\end{equation}
The last sum simplifies to $G(\lambda _{c}^{2}\Delta )S$, where 
\begin{equation}
G(x)=\frac{\sqrt{1-x^{2}}-1}{x^{2}}=-2\frac{\sin ^{2}(\frac{\sin ^{-1}x}{2})%
}{x^{2}},  \label{22}
\end{equation}
the second equality being a straightforward consequence of relations $\sqrt{%
1-x^{2}}=\cos \left( \sin ^{-1}x\right) $ and $\cos y-1=-2\sin ^{2}\frac{y}{2%
}$. Eventually, the continuity equation reads 
\begin{equation}
\frac{\partial R^{2}}{\partial t}+\vec{\nabla}\cdot \vec{j}_{RM}=0,
\label{23}
\end{equation}
where 
\begin{equation}
\vec{j}_{RM}=-\frac{2\hbar }{m}R^{2}\vec{\nabla}\left( G(\lambda
_{c}^{2}\Delta )S\right) .  \label{24}
\end{equation}
Alternatively, one can write the modified Schr\"{o}dinger equation as 
\begin{equation}
i\hbar \frac{\partial \Psi }{\partial t}=\frac{\hbar ^{2}}{2m}\left[ -\Delta
-\frac{i}{\rho }\left( \vec{\nabla}\cdot \vec{j}_{RM}\right) +\frac{i}{\rho }%
\left( \vec{\nabla}\cdot \vec{j}_{Sch}\right) \right] \Psi +V\Psi ,
\label{25}
\end{equation}
which is equivalent to 
\begin{equation}
i\hbar \frac{\partial \Psi }{\partial t}=-\frac{\hbar ^{2}}{2m}\left[ \Delta
+\frac{i}{\rho }\vec{\nabla}\cdot \left[ \rho \vec{\nabla}\left( 2G(\lambda
_{c}^{2}\Delta )-1\right) S\right] \right] \Psi +V\Psi .  \label{26}
\end{equation}
In general, the only notable exception to this rule being the modification
of Bia\l ynicki-Birula and Mycielski, the nonlinear modifications of the
Schr\"{o}dinger equation do not have the classical limit in the sense of the
Ehrenfest theorem. The discussed variant of the modification presented is
not exempted from this either. The nonlinear terms it introduces lead to
some corrections to the Ehrenfest relations. We will now work out these
corrections. For a general observable $A$ one finds that 
\begin{equation}
\frac{d}{dt}\left\langle A\right\rangle =\frac{d}{dt}\left\langle
A\right\rangle _{L}+\frac{d}{dt}\left\langle A\right\rangle _{NL},
\label{27}
\end{equation}
where the nonlinear contribution is due to $H_{NL}=H_{R}+iH_{I}$, $H_{R}$
and $H_{I}$ representing the real and imaginary part of $H_{NL}$,
respectively. The brackets $<>$ denote the mean value of the quantity
embraced. Specifying $A$ for the position and momentum operators one obtains
the general form of the modified Ehrenfest relations \cite{Pusz3} 
\begin{equation}
m\frac{d}{dt}\left\langle \vec{r}\right\rangle =\left\langle \vec{p}%
\right\rangle +I_{1},  \label{28}
\end{equation}
\begin{equation}
\frac{d}{dt}\left\langle \vec{p}\right\rangle =-\left\langle \vec{\nabla}%
V\right\rangle +I_{2},  \label{29}
\end{equation}
where 
\begin{equation}
I_{1}=\frac{2m}{\hbar }\int dV\vec{r}\rho H_{I},  \label{30}
\end{equation}
\begin{equation}
I_{2}=\int dV\rho \left( 2H_{I}\vec{\nabla}S-\vec{\nabla}H_{R}\right) .
\label{31}
\end{equation}
In the derivation of the last formula it was assumed that $\int dV\vec{\nabla%
}\left( \rho H_{I}\right) =0$ which for homogeneous modifications as the one
in question seems to be well justified \cite{Pusz3}. For the extension
concerned these integrals are found to be 
\begin{equation}
I_{1}=-\hbar \int dV\vec{r}\vec{\nabla}\cdot \left[ \rho \vec{\nabla}\left(
2G(\lambda _{c}^{2}\Delta )-1\right) S\right] ,  \label{32}
\end{equation}
\begin{equation}
I_{2}=-\frac{\hbar ^{2}}{m}\int dV\vec{\nabla}\cdot \left[ \rho \vec{\nabla}%
\left( 2G(\lambda _{c}^{2}\Delta )-1\right) S\right] \vec{\nabla}S.
\label{33}
\end{equation}

The presented modification is the best relativistic approximation to the
Schr\"{o}dinger equation that shares with it the properties of separability
for composed systems, the Galilean invariance, and the homogeneity in the
wave function. Moreover, the probability density of equation (25) (or (26))
is a non-negative function unlike that of the Klein-Gordon equation that
originates within the same framework. What distinguishes the
modification discussed from other nonlinear variants of the Schr\"{o}dinger
equation \cite{Bial, Doeb1, Pusz2} that encompass the same properties is
that they make additional assumptions or employ principles beyond those
specified here. These, essentially, are necessary to keep the number of new
terms at a reasonably manageable level. Such a problem does not arise here
as, in principle, there is only one new term, which certainly adds to the
attractiveness of this proposal. What is even more, our modification does
not introduce any new constants. This makes it truly unique in this respect
for all other nonlinear modifications of the Schr\"{o}dinger equation do
involve some new parameters.

The modification in question supports stationary states of the linear
Schr\"{o}dinger equation for which $S=-Et/\hbar +const$ without affecting
their energy $E$. Some other characteristic solutions to it include the
plane wave, but, interestingly enough, also ordinary Gaussian wave packets
and coherent states. Since both of the latter are the result of superposing
more elementary wave functions, each of which is a solution to the nonlinear
equation, this means that, similarly as the Schr\"{o}dinger equation, our
modification allows, although only partially, for the linear superposition
principle. It is easy to understand how this comes about here. The third
spatial derivative of the phase of a wave-packet, for simplicity in one
dimension ($\hbar =1$), 
\begin{equation}
S=\frac{mtx^{2}}{2\left( t^{2}+t_{0}^{2}\right) }-\frac{1}{2}\arctan \frac{t%
}{t_{0}},  \label{34}
\end{equation}
vanishes as do its higher order derivatives. Since the current correction of
(26) starts with $\vec{\nabla}^{3}S$ in the expansion, this means that as
far as  wave packets are concerned, the new term does not affect the
Schr\"{o}dinger equation at all. The same holds true for coherent states for
which $\Delta S=0$. The nonlinear modifications endowed with the discussed
property, that we chose to call the weak nonlinearity, are rare (see \cite
{Pusz2, Star} for the only other known examples of this kind).

As noted, the modification proposed employes only the most general and
universal assumptions. Because of that, it seems to be perfectly suited to
describe phenomena of some universal nature. As pointed out in \cite{Mens},
one of the most important phenomena in the quantum realm, the process of
decoherence responsible for rendering classical features of the ultimately
quantum world, can be modeled by Hamiltonians involving imaginary terms.
This is exactly the case of our modification. It introduces only such terms
and it does so in a minimal model-independent manner: the imaginary terms
that play the role of ``optical'' potentials emerge naturally, being an
integral part of the modification, and do not have to be put by hand. An
attractive consequence of this is that the process in question can be
thought of as caused by relativistic nonlinear corrections. One can hope
that the modification proposed will provide a good model to investigate this
process in greater detail.

\section*{Acknowledgments}

I would like to thank Professor Pawe{\l } O. Mazur for bringing to my
attention the work of Professor A. Staruszkiewicz that started my interest
in nonlinear modifications of the Schr\"{o}dinger equation and Professor G.
A. Goldin for his interest in this paper. A correspondence with Professor 
Wolfgang L\"{u}cke concerning the problem of separability in nonlinear quantum 
mechanics and a stimulating exchange of correspondence with Dr. Marek Czachor 
about many issues of nonlinear quantum mechanics are also gratefully 
acknowledged. This work was partially supported by the NSF grant No. 13020 F167 
and the ONR grant R\&T No. 3124141.

\end{document}